\documentstyle[12pt,psfig,draft,lscape]{nature-ejb}

\def\simlt{\mathrel{\hbox{\rlap{\hbox{\lower4pt\hbox{$\sim$}}}\hbox{$<$}}}}
\def\simgt{\mathrel{\hbox{\rlap{\hbox{\lower4pt\hbox{$\sim$}}}\hbox{$>$}}}}

\def\ale{\mathrel{\hbox{\rlap{\hbox{\lower4pt\hbox{$\sim$}}}\hbox{$<$}}}}
\def\age{\mathrel{\hbox{\rlap{\hbox{\lower4pt\hbox{$\sim$}}}\hbox{$>$}}}}

\def\grb{GRB~030329}
\newcommand{\hete}{\textit{HETE-II}}

\def\spose#1{\hbox to 0pt{#1\hss}}
\newcommand\lsim{\mathrel{\spose{\lower 3pt\hbox{$\mathchar"218$}}
     \raise 2.0pt\hbox{$\mathchar"13C$}}}
\newcommand\gsim{\mathrel{\spose{\lower 3pt\hbox{$\mathchar"218$}}
     \raise 2.0pt\hbox{$\mathchar"13E$}}}

\begin{document}


\title{\Large \bf
Discovery of the Bright Afterglow of the Nearby Gamma-Ray Burst
of 29 March 2003}

\author{
   P.~A.~Price\affiliation[1]
      {RSAA, ANU, Mt.\ Stromlo Observatory,
       via Cotter Rd, Weston Creek, ACT, 2611, Australia},
   D.~W.~Fox\affiliation[2]
   {Caltech Optical Observatories 105-24, California Institute of
    Technology, Pasadena, CA\,91125, USA},
   S.~R.~Kulkarni\affiliationmark[2],
   B.~A.~Peterson\affiliationmark[1],
   B.~P.~Schmidt\affiliationmark[1],
   A.~M.~Soderberg\affiliationmark[2],
   S.~A.~Yost\affiliation[3]
    {Space Radiation Laboratory 220-47, Caltech,
     Pasadena, CA\,91125, USA},
   E.~Berger\affiliationmark[2],
   S.~G.~Djorgovski\affiliationmark[2],
   D.~A.~Frail\affiliation[4]
    {National Radio Astronomy Observatory, P.O. Box 0, Socorro, New
     Mexico 87801, USA},
   F.~A.~Harrison\affiliationmark[3],
   R.~Sari\affiliation[5]
    {Theoretical Astrophysics 130-33, 
     Caltech, Pasadena, CA\,91125, USA},
   A.~W.~Blain\affiliationmark[2],
   and
   S.~C.~Chapman\affiliationmark[2].
}



\date{\today}{} \headertitle{Discovery of Nearby GRB}
\mainauthor{Price et al.}

\summary {

Many past studies of cosmological $\gamma$-ray bursts (GRBs) have been
limited because of the large distance to typical GRBs, resulting in
faint afterglows.  There has long been a recognition that a nearby GRB
would shed light on the origin of these mysterious cosmic explosions,
as well as the physics of their fireballs.  However, GRBs nearer than
$z=0.2$ are extremely rare, with an estimated rate of localisation of
one every decade\cite{s99}.  Here, we report the discovery of bright
optical afterglow emission from \grb\cite{vcd+03}.  Our prompt
dissemination\cite{pp03} and the brilliance of the afterglow resulted
in extensive followup (more than 65 telescopes) from radio through
X-ray bands, as well as measurement of the redshift,
$z=0.169$\cite{gpe+03}.  The $\gamma$-ray and afterglow properties of
\grb\ are similar to those of cosmological GRBs (after accounting for
the small distance), making this the nearest known cosmological GRB.
Observations have already securely identified the progenitor as a
massive star that exploded as a supernova\cite{smg+03}, and we
anticipate futher revelations of the GRB phenomenon from studies of
this source.

}

\maketitle


On 2003 March 29, 11$^{\rm h}$37$^{\rm m}$14$^{\rm s}$.67 UT, \grb\
triggered all three instruments on board the High Energy Transient
Explorer II (\hete). About 1.4~hours later, the \hete\ team
disseminated via the GRB Coordinates Network (GCN) the 4-arcminute
diameter localisation\cite{vcd+03} by the Soft X-ray Camera (SXC).  We
immediately observed the error-circle with the Wide Field Imager (WFI)
on the 40-inch telescope at Siding Spring Observatory under inclement
conditions (nearby thunderstorms).  Nevertheless, we were able to
identify a bright source not present on the Digitised Sky Survey
(Figure~\ref{fig:discovery}) and rapidly communicated the discovery to
the community \cite{pp03}.  The same source was independently detected
by the RIKEN automated telescope\cite{t03}.

With a magnitude of 12.6 in the $R$ band at 1.5~hours, the optical
afterglow of \grb\ is unusually bright. At the same epoch, the
well-studied GRB~021004 was $R \sim 16\,$mag\cite{fyk+03}, and the
famous GRB~990123 was $R\sim 17\,$mag\cite{abb+99}.  The brightness of
this afterglow triggered observations by over 65 optical telescopes
around the world, ranging from sub-metre apertures to the Keck~I
10-metre telescope.  Unprecedented bright emission at
radio\cite{bsf03}, millimetre\cite{ksn03},
sub-millimetre\cite{hmt+03}, and X-ray\cite{ms03} was also reported
(Figure~\ref{fig:broadband}).

Greiner et al.\cite{gpe+03} made spectroscopic observations with the
Very Large Telescope (VLT) in Chile approximately 16~hours after the
GRB and, based on absorption as well as emission lines, announced a
redshift of $z=0.1685$.  From Keck spectroscopic observations obtained
8~hours later (Figure~\ref{fig:spectrum}) we confirm the VLT redshift,
finding $z=0.169\pm 0.001$.  We note that the optical afterglow of
\grb\ was, at 1.5~hours, approximately the same brightness as the
nearest quasar, 3C~273 ($z=0.158$); it is remarkable that such a large
difference in the mass of the engine can produce an optical source
with the same luminosity.

%

With a duration of about 25-s and multi-pulse profile\cite{vcd+03},
\grb\ is typical of the long duration class of GRBs.  The fluence of
\grb, as detected by the Konus experiment \cite{gmp+03}, of
$1.6\times10^{-4}$~erg~cm$^{-2}$ (in the energy range 15-5000~keV)
places this burst in the top 1\%\ of GRBs.

At a redshift of 0.169, \grb\ is the nearest of the cosmological GRBs
studied in the 6-year history of afterglow research.  Assuming a
Lambda cosmology with $H_0 = 71$~km/s/Mpc, $\Omega_M = 0.27$ and
$\Omega_\Lambda=0.73$, the angular-diameter distance is $d_A=589\,$Mpc
and the luminosity distance is $d_L=805\,$Mpc.  The isotropic
$\gamma$-ray energy release, $E_{\gamma,\rm iso} \sim 1.3\times
10^{52}\,$erg, is typical of cosmological GRBs\cite{fks+01}.
Likewise, the optical and radio luminosities of the afterglow of \grb\
are not markedly different from those of cosmological GRBs.  In
particular, the extrapolated isotropic X-ray luminosity at $t=10$ hr
is $L_{X,\rm iso} \sim 6.4\times 10^{45}\,$erg\,s$^{-1}$, not
distinctly different from that of other X-ray afterglows (e.g. ref.\
\pcite{bkf03}).

Nonetheless, two peculiarities about the afterglow of \grb\ are worth
noting.  First, the optical emission steepens from $f(t) \propto
t^{-\alpha}$ with $\alpha_1 = 0.873 \pm 0.025$ to $\alpha_2 = 1.97 \pm
0.12$ at epoch $t_* = 0.481 \pm 0.033\,$d
(Figure~\ref{fig:lightcurve}; see also ref.\ \pcite{gsb03}).  This
change in $\alpha$ is too large to be due to the passage of a cooling
break ($\Delta\alpha=1/4$; ref.\ \pcite{spn98}) through the optical
bands.  On the other hand, $\alpha\sim 2$ is typically seen in
afterglows following the so-called ``jet-break'' epoch ($t_j$). Before
this epoch, the explosion can be regarded as isotropic, and following
this epoch the true collimated geometry is manifested.  Such an early
jet break would imply a substantially-lower energy release than
$E_{\gamma,\rm iso}$.

If $t_*\sim t_j$ then using the formalism and adopting the density and
$\gamma$-ray efficiency normalisations of Frail et al.\cite{fks+01} we
estimate the true $\gamma$-ray energy release to be $E_\gamma \sim
3\times 10^{49}$~erg.  This estimate is $4\sigma$ lower than the
``standard energy'' of $5\times 10^{50}$~erg found by Frail et
al.\cite{fks+01} The geometric-corrected x-ray luminosity\cite{bkf03}
would also be the lowest of all x-ray afterglows.  If the above
interpretation is correct then \grb\ may be the missing link between
cosmological GRBs and the peculiar GRB~980425\cite{gvv+98}
($E_{\gamma,\rm iso}\sim 10^{48}\,$erg) which has been associated with
SN~1998bw at $z=0.0085$.

Second, the decay of the optical afterglow is marked by bumps and
wiggles (e.g.\ ref.\ \pcite{log03}).  These features could be due to
inhomogeneities in the circumburst medium or additional shells of
ejecta from the central engine. In either case, the bumps and wiggles
complicate the simple jet interpretation offered above.  We note that
if the GRB had been more distant, and hence the afterglow fainter,
then the break in the light curve would likely have been interpreted
as the jet break without question.

The proximity of \grb\ offers us several new opportunities to
understand the origin of GRBs.  Red bumps in the light curve have been
seen in several more-distant ($z\sim 0.3$ to 1) GRB afterglows (e.g.\
refs \pcite{bkd+99},\pcite{gsw+03}) and interpreted as underlying SNe
that caused the GRBs.  While these red bumps appeared to be consistent
with a SN light curve, prior to \grb, it had not yet been
unambiguously demonstrated that they were indeed SNe.  As this paper
was being written, a clear spectroscopic signature for an underlying
SN has been identified in the optical afterglow of \grb\cite{smg+03}.
Our own spectroscopy at Palomar and Keck confirm the presence of these
SN features.  This demonstrates once and for all that the progenitors
of at least some GRBs are massive stars that explode as SNe.

However, there remain a number of issues still to be resolved,
relating to the physics of GRB afterglows and the environment around
the progenitor.  The ``fireball'' model of GRB afterglows predicts a
broad-band spectrum from centimetre wavelengths to x-rays that evolves
as the GRB ejecta expand and sweep up the surrounding
medium\cite{spn98}.  Testing this model in detail has in the past been
difficult, primarily due to both low signal-to-noise and interstellar
scintillation at the longer wavelengths, from which come the majority
of spectral and temporal coverage of the afterglow evolution (e.g.\
refs. \pcite{bsf+00}, \pcite{pk01}).  \grb, with bright emission at
all wavelengths (Figure~\ref{fig:broadband}), and limited
scintillation (due to the larger apparent size) will allow astronomers
to test the predictions of the fireball model with unprecedented
precision through the time evolution of the broad-band spectrum, the
angular size of the fireball, and its proper motion (if any).  It has
been long predicted that if the progenitors of GRBs are massive stars,
the circumburst medium should be rich and inhomogenous\cite{cl99}, but
it has been difficult to find evidence for this.  However, for \grb,
it should be possible to trace the distribution of circumburst
material and determine the environment of the progenitor.

Even in the Swift (launch December 2003) era, we expect only one such
nearby ($z < 0.2$) GRB every decade (scaling from ref.\ \pcite{s99}).
Thus \grb, the nearest of the cosmological GRBs to date, has given
astronomers a rare opportunity to be up close and personal with a GRB
and its afterglow.  We eagerly await reports of the many experiments
that have been and will be conducted to shed new light on the GRB
phenomenon.

\section*{Acknowledgments} 

PAP and BPS thank the ARC for supporting Australian GRB research.  GRB
research at Caltech is supported in part by funds from NSF and NASA.
We are, as always, indebted to Scott Barthelmy and the GCN, as well as
the \hete\ team for prompt alerts of GRB localisations.




\clearpage

\begin{figure}
\centerline{\psfig{file=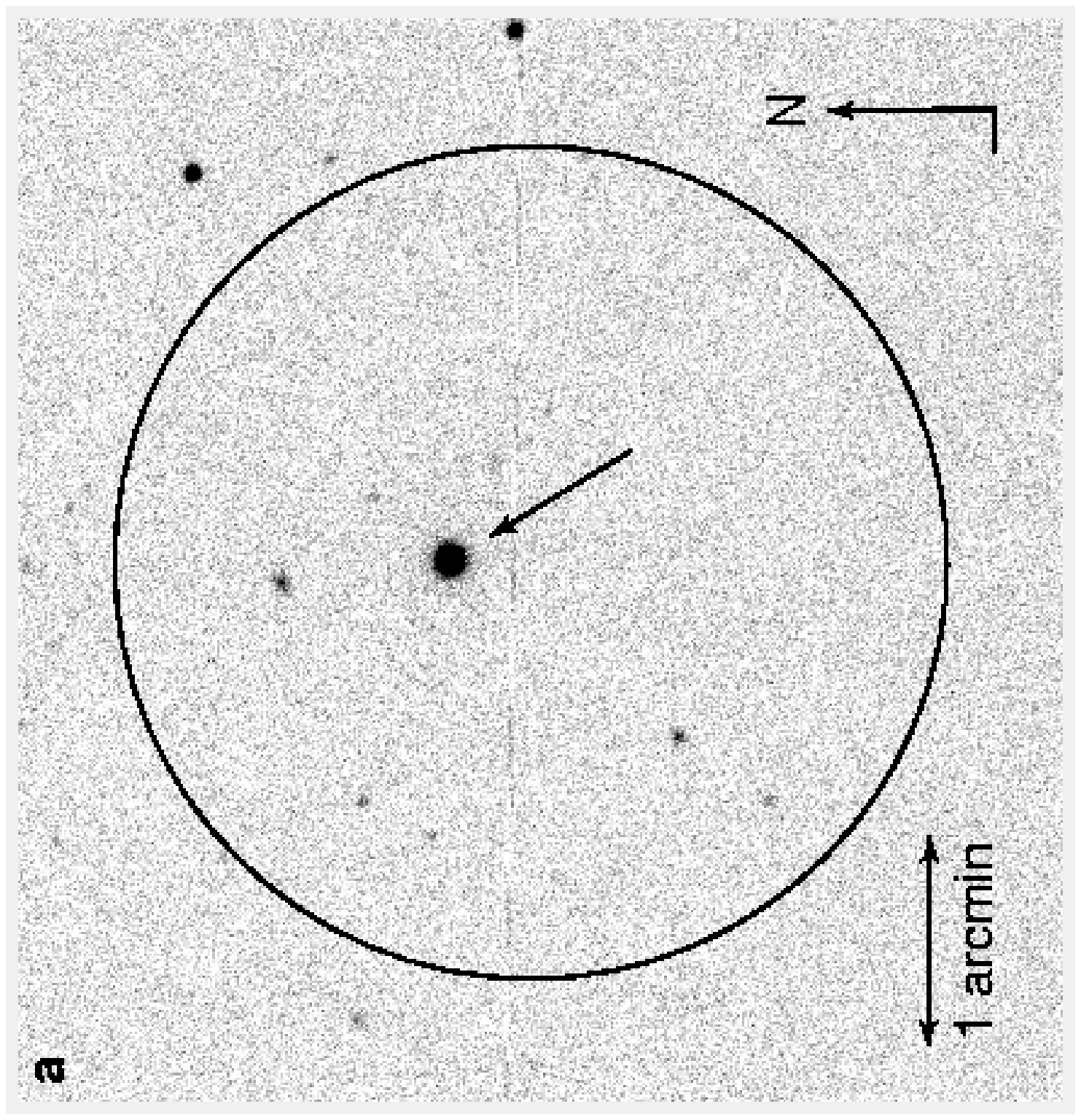,width=4in}}
\centerline{\psfig{file=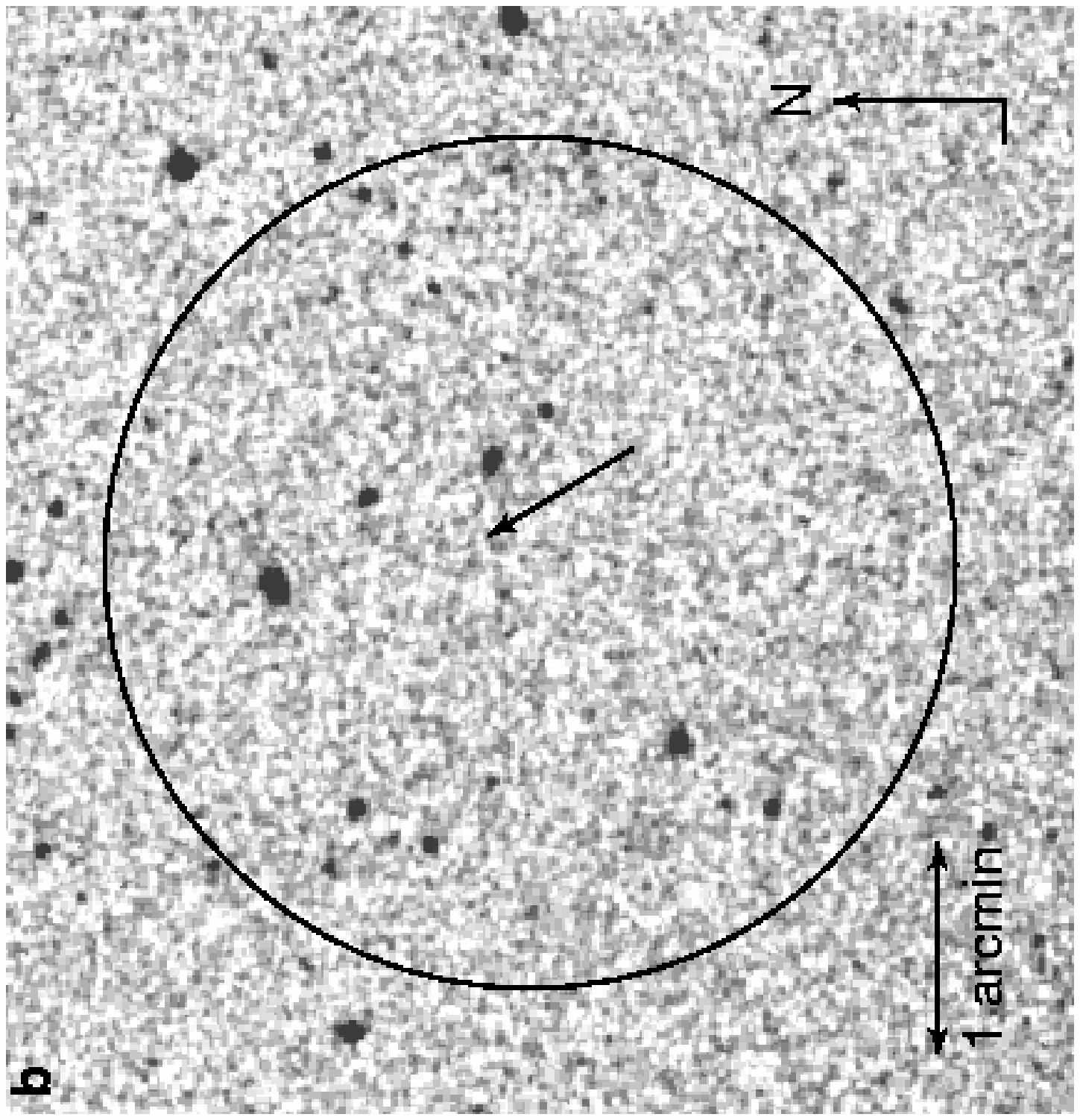,width=4in}}
\caption[]{}
\label{fig:discovery}
\end{figure}

\noindent {\bf Figure \ref{fig:discovery}:} Discovery of the bright optical afterglow of
\grb. The 600-s exposure taken with the Wide-Field Imager at the
40-inch telescope of the Siding Spring Observatory (SSO) using an
$R$-filter (a) started at March 29, 13$^{\rm h}$05$^{\rm m}$ UT,
2003, about 1.5 hours after the $\gamma$-ray event, and was strongly
affected by clouds.  Nevertheless, comparison of the SSO image with
the Second Digitised Sky Survey (b) in the $R_F$-band at the
telescope allowed us to identify a 12th magnitude afterglow (arrowed)
within the 4-arcmin \hete\ SXC error-circle\cite{vcd+03} (marked). The
position of the optical afterglow was determined with respect to
USNO-A2.0 and found to be $\alpha_{2000} = 10^{\rm h}44^{\rm m}59^{\rm
s}.95$, $ \delta_{2000} = +21^\circ31^\prime17^{\prime\prime}.8$ with
uncertainty of 0.5~arcsec in each axis.

\begin{figure}
\centerline{\psfig{file=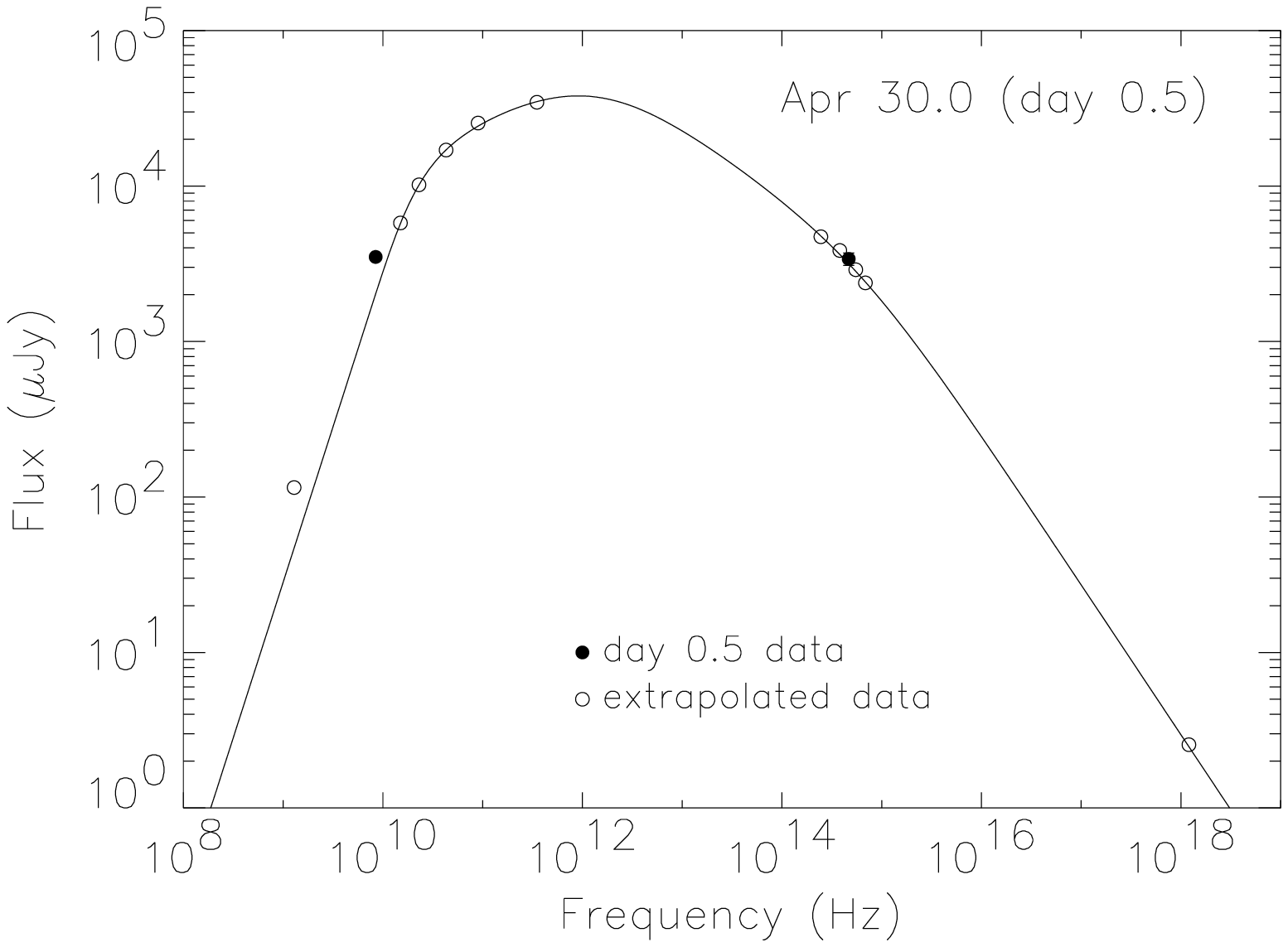,width=6in}}
\caption[]{}
\label{fig:broadband}
\end{figure}

\noindent {\bf Figure \ref{fig:broadband}:} A snapshot spectral flux
distribution of the afterglow of GRB~030329.  This broad-band spectrum
of the afterglow at 0.5~days after the
GRB\cite{ksn03,hmt+03,ms03,rcb03,p03,bsp+03,zbt+03} demonstrates both
the brightness of the afterglow, with the resulting spectral coverage.
Solid circles represent measurements made near the nominal time; open
circles represent measurements extrapolated to the nominal time
assuming evolution appropriate for a constant-density medium; this
figure is therefore meant to be illustrative rather than entirely
accurate.  A simple fit of an afterglow broad-band spectrum yields the
following spectral parameters (we use the convention and symbols of
ref.\ \pcite{spn98}): synchrotron self-absorption frequency, $\nu_a
\sim 25\,$GHz; peak frequency, $\nu_m\sim 1270\,$GHz; cooling
frequency, $\nu_c\sim 6.2\times 10^{14}\,$Hz; peak flux, $f_m\sim
65\,$mJy; and electron energy index, $p\sim 2$.  The physical
parameters inferred are as follows: explosion energy, $E\sim 5.7\times
10^{51}\,$erg; ambient density, $n\sim 5.5$ atom cm$^{-3}$; electron
energy fraction, $\epsilon_e\sim 0.16$ and magnetic energy fraction,
$\epsilon_B\sim 0.012$.

\begin{figure}
\centerline{\psfig{file=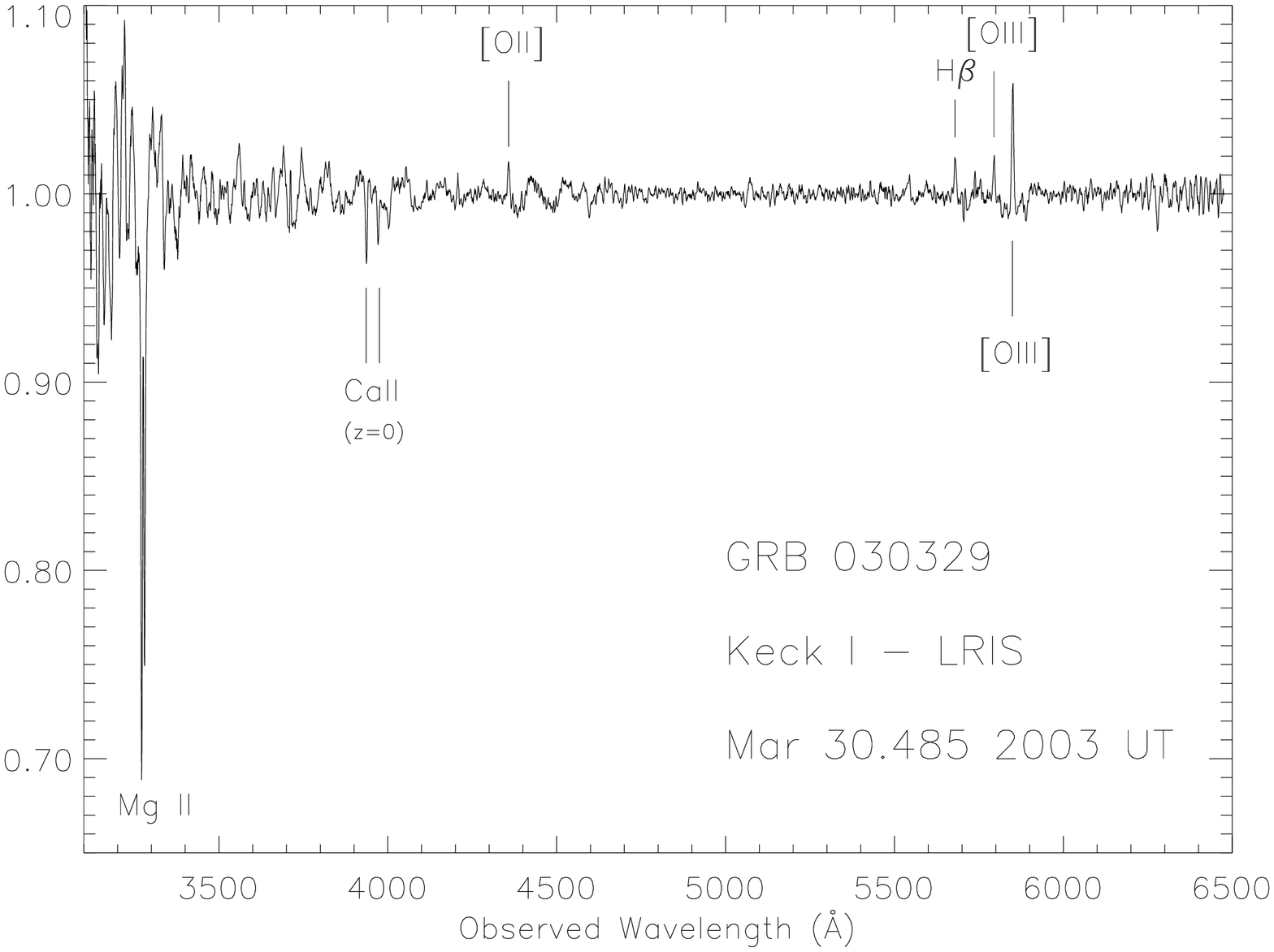,width=6in,angle=90}}
\caption[]{}
 
\label{fig:spectrum}
\end{figure}

\noindent {\bf Figure \ref{fig:spectrum}:} Spectrum of the optical
afterglow.  The observation was made with the Low Resolution Imaging
Spectrometer on the Keck I telescope, using the 400~lines/mm grism on
the blue side, giving an effective resolution of 4.2\AA.  Our
observation consisted of a single 600~second exposure on the
afterglow.  We reduced and extracted the spectra in the standard
manner using IRAF.  No standard star observations were available, so
we have simply fit and normalised the continuum before smoothing with
a 4\AA\ boxcar.  We identify narrow emission lines from [O~II],
H$\beta$ and [O~III], and absorption lines from Mg~II at a mean
redshift, $z=0.169 \pm 0.001$, making GRB~030329 the lowest-redshift
cosmological GRB.  These emission lines are typical of star-forming
galaxies, whereas the absorption lines are caused by gas in the disk
of the galaxy.  In addition, we identify Ca~II at $z\approx 0$,
presumably due to clouds in our own Galaxy.

\begin{figure}
\centerline{\psfig{file=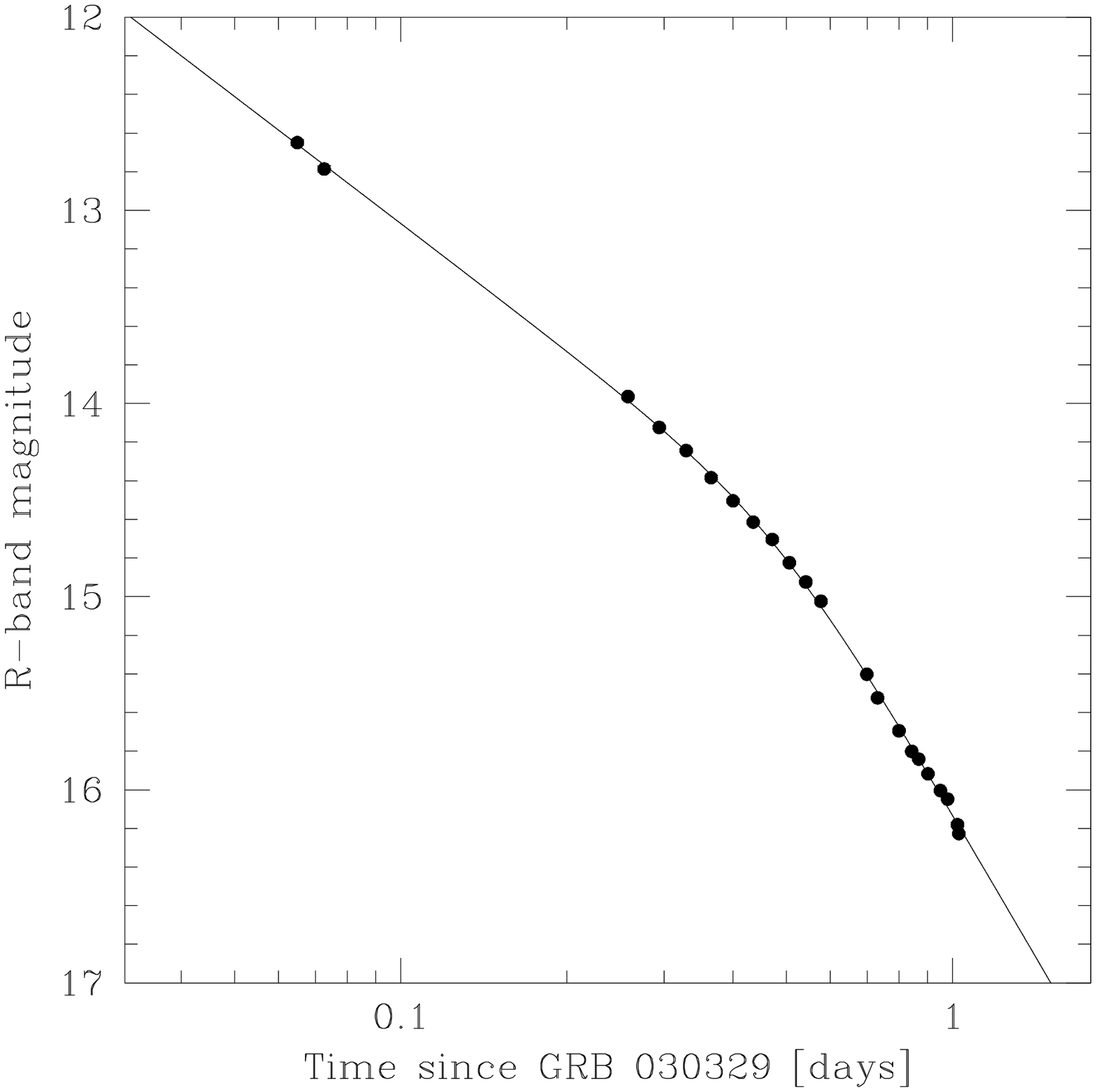,width=6in,angle=0}}
\caption[]{}
\label{fig:lightcurve}
\end{figure}

\noindent {\bf Figure \ref{fig:lightcurve}:} Light-curve of the
optical afterglow of \grb.  This $R$-band light curve spans from our
discovery to approximately 1~day after the GRB.  Due to the brightness
of the GRB, errors in the measurements are smaller than the plotted
points.  In addition to measurements gleaned from the GCN
Circulars\cite{bsp+03,fo03}, we include the following measurements
from observations with the SSO 40-inch telescope with WFI in 2003
March: 29.5491, $R = 12.649 \pm 0.015$~mag; 29.5568, $R = 12.786 \pm
0.017$~mag; 30.5044, $R = 16.181 \pm 0.010$~mag; 30.5100, $R = 16.227
\pm 0.009$~mag.  These measurements are relative to field stars
calibrated by Henden\cite{h03}.

\clearpage

\clearpage

\clearpage

\clearpage

\end{document}